\documentclass[nofootinbib,reprint,amsmath,amssymb,aps]{revtex4-1}
\usepackage[utf8]{inputenc}
\usepackage{graphicx}
\usepackage[dvipsnames]{xcolor}
\usepackage{dcolumn}
\usepackage{amssymb}
\usepackage{comment}

\usepackage{mathabx}
\usepackage{mathrsfs}  
\usepackage{tensor}
\usepackage[normalem]{ulem}
\usepackage{natbib}
\usepackage{cancel}
\usepackage{bm}
\newcommand{\qm}[1]{``#1''}
\usepackage{hyperref}
\usepackage{stackengine,scalerel}
\hypersetup{colorlinks, linkcolor={red},citecolor={blue},urlcolor={blue}}

\newcommand{\dd}{{\rm d}}

\begin{document}

\title[Compact binary system dynamics at the second post-Newtonian order]{Compact binary system dynamics at the second post-Newtonian order:\\
analytical formula of the coordinate time for eccentric and circular orbits}

\author{Vittorio De Falco$^{1}$}\email{deltafi.mat@live.it}
\author{Marco Gallo$^{1}$} \email{gallomarco1989@gmail.com}

\affiliation{
$^1$ Ministero dell'Istruzione e del Merito (M.I.M., ex M.I.U.R.)
}

\date{\today}

\begin{abstract}
This work is based on the letter \emph{Phys. Lett. B, 865,
139484 (2025)}, where we developed the analytical expression of the coordinate time in terms of the eccentric anomaly at the second post-Newtonian order in General Relativity for a compact binary system moving on eccentric orbits. The aim of this paper is to provide more details about the performed calculations and to produce other new results. More specifically, we will focus on deriving the analytical expression of the coordinate time at the second Post-Newtonian order for circular orbits and then discuss two astrophysical applications involving binary neutron star and black hole systems.
\end{abstract}
\maketitle

\section{Introduction}
General Relativity (GR) is experiencing a flourishing epoch thanks to some breakthrough observational probes that have occurred in astrophysics in the recent years, like the direct detection of gravitational waves (GWs) from the merging of two black holes (BHs) \cite{Abbott2016} and from two neutron stars (NSs) \cite{Abbott2017-NS}, as announced by the Virgo and LIGO collaborations in 2015; in 2019 the first image of the supermassive BH located at the centre of the M87 galaxy was released by the Event Horizon Telescope collaboration \cite{EHC20191}. However, other fundamental probes supporting GR have been confirmed over the years  \cite{Clifford2014,Ishak2019,Volkel2021}. 

These promising events do not make scientists stop inquiring into gravity, on the contrary they encourage them to provide new solid tests. From an astrophysical perspective, gravity is still not fully understood in the strong field regime and other alternatives must be considered. In this direction, one of the main targets of investigation is represented by the class of compact binary systems. 

From a theoretical point of view, these systems are very difficult to be described, because the nonlinear structure of GR makes even complicate to derive the equations of motion (EoMs) \cite{Blanchet2014}. However, even if we were able to determine them, they would be governed by retarded-partial-integro differential equations \cite{Maggiore:GWs_Vol1,Blanchet2014,Poisson-Will2014}, as the compact objects are considered to be continuous bodies.

A practical way to solve this issue and, at the same time, to save the observational techniques developed in classical astronomy is to resort to the \emph{post-Newtonian (PN) approximation method} \cite{Lorentz1937}. This approach is based on the following two crucial assumptions \cite{Maggiore:GWs_Vol1,Blanchet2014,Poisson-Will2014}: (1) the two bodies are slowly moving, weakly self-gravitating, and weakly stressed (also known as \emph{PN gravitational sources}); (2) the two objects are well separated. In these hypotheses, we can first consider the bodies as two test particles and the curved background can be approximated as a Newtonian absolute Euclidean space on which we can add the relativistic corrections in power-series of $1/c$ \cite{Blanchet2014}. In particular, the $n$PN order indicates to expand a formula up to order $1/c^{2n}$ and neglect higher terms. The practical effect of this strategy consists of dealing with ordinary differential equations, which preserve their relativistic nature (i.e., invariance under a global PN-expanded Lorentz transformation) \cite{Blanchet2014,Poisson-Will2014}. Instead, they lose the GR general covariance, as particular coordinate systems are employed (e.g., harmonic coordinates).  

The PN technique is widely and successfully exploited in the astrophysical literature \cite{Blanchet2024}, where the various fields of application are: gravitational back-reaction in compact binary systems \cite{Taylor:1989sw,Damour:1990wz,Kramer:2006nb,Kramer:2021jcw}; direct detection of GWs from coalescing compact binaries \cite{Blanchet2014}; precision tests of gravity theories \cite{Wex:2014abc}; NS mass measurements in binary pulsars \cite{Ozel:2016oaf}; inquiring the extreme mass ratio inspiral candidates and looking for their nature \cite{Seoane2018}.

The focus of this work is on BH binaries and relative emission of GWs, which are closely related and where the PN method is extensively employed. The state of the art on the conservative dynamics for nonspinning compact binaries is derived at the 4PN order following different formalisms \cite{Damour2014j,Damour2016a,Bernard2015,Marchand2017,Foffa2019a,Foffa2019b} and has been also extended to the 5PN and 6PN levels, but admitting some unknown variables to be determined \cite{Bini2019a,Bini2020a,Bini2020c,Bini2020r}. The GW templates, instead, have been derived up to the 4PN accuracy \cite{Marchand2020,Larrouturou2021a,Larrouturou2021b}. 

Finally, for spinning compact binaries, the effects in the radiative field and the energy flux are such that  the spin-orbit (SO) contribution is known at 4PN \cite{Bohe2013c,Marsat2013c}, whereas the spin-spin (SS) \cite{Bohe2015a,Cho2021} and the spin-spin-spin (SSS) \cite{Marsat2014} at 3PN; the SO, SS, and SSS interactions in the EoMs have been computed at 3.5PN   \cite{Damour2007a,Marsat2012,Bohe2012,Levi_2016}, 3PN \cite{Bohe2015a,Levi2014,Cho2021}, and 3.5PN level \cite{Marsat2014}, respectively.

After the first GW detection occurred in 2015, more than one hundred of GW events have been observed from astrophysical compact binary mergers (examples are: BH-BH, NS-NS, and looking for BH-NS candidates) \cite{Abbott2022,Soni2024}. In addition, LIGO, Virgo, and KAGRA' sensitivities have been updated such that the GW detection rate has been powered to be four times broader than the previous one, leading thus to the current fourth observational run \cite{Cahillane2022,Soni2024}. In this unstoppable observational GW detection survey, there will be other additional and complementary participants, represented by the future third-generation ground-based GW interferometers Einstein Telescope \cite{ET_SCIENCE2020JCAP} and Cosmic Explorer \cite{Reitze2019a,Reitze2019b}, as well as the space-based devices LISA \cite{Audley2017} and TianQin \cite{TianQin2015}.

This unquenchable attention towards an increase in our observational capacity to investigate gravity and inspiralling BH binaries must be accompanied by appropriate theoretical assessments. This justifies why there is a great ferment in achieving higher and higher PN orders. This is extremely helpful both for extracting information on the gravitational source under study \cite{Flanagan1998,Mingarelli2019,Bernard2019,Soma2023}, but also for generating template matching in ground- and space-based GW astronomy to fit the observational data \cite{Schmidt2024}. 

It is important to underline that the GW signal $h(T)$ is a parametric plot having on the $x$-axis the coordinate time $T$ related to the BH binary's dynamics and on the $y$-axis the strain $h(T)$ measured in the detector frame. Both functions $h(T)$ and $T$ depend on an angle parametrizing the motion. Generally, the coordinate time is computed numerically, making thus time-consuming the production of the GW templates and therefore arduous the fit of the observational data. Identically, for pulsar timing the pulse profile (depending on the coordinate time $T$) is computed numerically leading thus to develop suitable and elaborate algorithms to speed up the ensuing calculations.

The aim of this paper is to derive the analytical expression of the coordinate time $T$ at the 2PN level as a function of the eccentric anomaly $u$. Therefore, providing $T(u)$, it allows us to have $h(T(u))$ as an analytical function, making thus easier and faster the production of GW templates, as well as the benchmark against the observations. The same reasoning also applies to pulsar timing, as it drastically reduces the implementation of numerical schemes within the software like \texttt{TEMPO} \cite{Taylor:1989sw}, \texttt{TEMPO2} \cite{Edwards:2006zg}, and \texttt{PINT} \cite{Luo:2020ksx}. Moreover, it is applicable to coherent pulsar search algorithms \cite{lentali2018,freire2018}. This approach demonstrates substantial efficacy even at low PN orders, delivering tangible practical outcomes. The importance of this work lies primarily in the introduction of a novel methodology, absent from previous literature, which can be readily extended to higher PN orders.

This work provides more details about the derivation of the analytical expression of the coordinate time $T$ at the 2PN level, already obtained in the letter \cite{Defalco2024}. In this view, it is useful to recall that in a previous work \cite{LetteraDBA}, De Falco and collaborators have derived the 1PN-accurate analytical formula of $T(\varphi)$, where $\varphi$ is the polar angle in the plane where the binary dynamics occurs. This result is obtained by considering the analytical formula of the relative distance $R$ between the two bodies as function of $\varphi$, namely $R(\varphi)$, provided by Damour and Deruelle \cite{Damour1985}. The derivation of $T(\varphi)$, using the $\varphi$-parametrization, is composed of three steps: (1) determining the differential equation $\dd T=f(\varphi)\dd\varphi$, where $f(\varphi)$ is the integrating function expressed up to the 1PN level; (2) analytically integrating the function $f(\varphi)$; (3) the final expression gives rise to a non-continuous function, as it shows periodic non-connected branches, not adapted to describe the coordinate time (being a monotonically increasing function). The cause of this effect is the presence of the tangent function, appearing in a nonlinear way in the ensuing integrated formula. Therefore, to attain the continuity, we introduce the concept of \emph{accumulation function}, which at every period smoothly connects the periodic curves for having then a regular map. The final result perfectly matches the coordinate time computed numerically. 

In this paper, we show how to determine $T(u)$ at the 2PN order as a function of the eccentric anomaly $u$. This time, the alternative $u$-parametrization is due to the different orbit description offered by Sch\"afer and Wex at the 2PN order \cite{SchaferWex-1993}. It is important to note that the authors do not provide an analytical expression of $R(\varphi)$, as this expression is extremely demanding due to the involved equations, which turns out to be highly nonlinear. Our analytical expression is the combination of various intertwined procedures, which will be better explained in this article. The surprising result is that up to the 1PN level, we do not need to invoke the accumulation function, whereas it must be necessarily exploited to work out the 2PN term. BH binary's dynamics and related GW emission are breeding grounds for applying the 2PN coordinate time formula, where higher PN schemes are fundamental. Instead, for pulsar binary systems the 2PN contribution seems to be less essential\footnote{It seems that there is no implementation of the 2PN orbit in any of the existing timing packages (e.g., \texttt{TEMPO2} and \texttt{PINT}), since there is no known system so far where the Sch{\"a}fer \& Wex 2PN terms \cite{SchaferWex-1993} are even remotely relevant (N. Wex, priv. comm.).}. However, through our analytical formula it is easier to fit the observational data and thus the extraction of the 2PN parameters related to the binary compact object under study. Indeed, having the two bodies' masses, the relative distance, the relative radial velocity, and the orbital eccentricity we are able to estimate the relativistic precession and reconstruct the orbital motion up to the 2PN order.

The paper is organized as follows: in Sect. \ref{sec:setting} we provide the physics and mathematical setting related to the motion of compact binary systems; we first focus on the elliptic dynamics (see Sect. \ref{sec:elliptic}) and then on the circular orbits (see Sect. \ref{sec:circular}); we then provide two astrophysical applications of our formula (see Sect. \ref{sec:application}); we finally draw the conclusions and future perspectives (see Sect. \ref{sec:end}).  

\section{Compact binary system setting}
\label{sec:setting}
We consider a compact binary system composed of two self-gravitating bodies with masses $m_1>m_2$, total mass $M=m_1+m_2$, position vectors $\bm{r_1}(T)$ and $\bm{r_2}(T)$ and, velocity vectors $\bm{v_1}(T)$ and $\bm{v_2}(T)$, where $T$ is the coordinate time.  Moreover, we define the separation vector $\bm{R}(T)=\bm{r_1}(T)-\bm{r_2}(T)$ and its modulus $R=|\bm{R}(T)|$, the reduced mass $\mu=m_1 m_2/M$, the symmetric mass ratio $\nu=\mu/M$, and the unit vector $\bm{n}=\bm{R}(T)/R$. Introducing $\bm{r}=\bm{R}/(GM)$ and $r=|\bm{r}|$, the coordinate time scales as $t=T(GM)$ \cite{Memmesheimer-2004}. The EoMs at the 2PN order in the center-of-mass frame in harmonic coordinates are derived via the Hamiltonian formalism and read as \cite{Memmesheimer-2004,SchaferWex-1993}: 
\begin{subequations} \label{eq:EoMs}
\begin{align}
{\dot r}^2&\equiv\frac{1}{s^4}\left( \frac{\dd s}{\dd t} \right)^2=\sum_{j=0}^{5} A_j s^j,\label{eq:EoM1}\\
\dfrac{\dd\varphi}{\dd s}&=\dfrac{1}{\sqrt{(s_{-}-s)(s-s_{+})}}\sum_{j=0}^{3} B_j s^j,\label{eq:EoM2}
\end{align}    
\end{subequations}
where $s=1/r$, $s_{-}$ and $s_{+}$ are the inverse of the apastron and periastron, respectively\footnote{It is important to note that at the 0PN level, we have
$$
s_{-}=\frac{1+e_0}{h_0^2}=\frac{(GM)^{-1}}{a(1-e_0)},\qquad s_{+}=\frac{1-e_0}{h_0^2}=\frac{(GM)^{-1}}{a(1+e_0)},
$$
where the lengths are scaled by $GM$. We have used the classical relation $J_0=GMa(1-e_0^2)$, where $e_0$ is the 0PN eccentricity.}. The last quantities are determined by searching for the non-vanishing positive roots having finite limit as $\frac{1}{c}\rightarrow 0$ of the fifth-degree polynomial in Eq. \eqref{eq:EoM1}. The coefficients $A_j$ (see Eq. (A3a) in Ref. \cite{Memmesheimer-2004}) and $B_j$ (see Eq. (A4a) in Ref. \cite{Memmesheimer-2004}) are functions of $\nu$, 2PN energy $E=E_0+\frac{1}{c^2}E_1+\frac{1}{c^4}E_2+\mathcal{O}(c^{-6})$, and reduced 2PN angular momentum $h=h_0+\frac{1}{c^2}h_1+\frac{1}{c^4}h_2+\mathcal{O}(c^{-6})$ with $h=J/(GM)$ and $J$ being the angular momentum. The explicit expressions of $E$ and $h$ can be found in Ref. \cite{Memmesheimer-2004}. The EoMs can be written by employing the following \qm{Keplerian-like} parametrization \cite{Memmesheimer-2004}:
\begin{subequations} \label{eq:QuasiKepPar}
\begin{align}
r&=a_r(1-e_r \cos{u}),\label{eq:QuasiKepPar1}\\
l&=u-e_t \sin{u}+\frac{g_{4t}}{c^4}(v-u)+\frac{f_{4t}}{c^4}\sin{v},\label{eq:QuasiKepPar2}\\
\frac{2\pi}{K}(\varphi-\varphi_0)&=v+\frac{f_{4\phi}}{c^4}\sin{(2v)}+\frac{g_{4\phi}}{c^4}\sin{(3v)},\label{eq:QuasiKepPar3}\\
v&=2\arctan{\left[\left(\frac{1+e_{\varphi}}{1-e_{\varphi}}\right)^{1/2}\tan{\frac{u}{2}}\right]},\label{eq:QuasiKepPar4}
\end{align}    
\end{subequations}
where $u$ and $v$ are the eccentric and true anomalies, respectively, $\varphi$ the polar angle, $l=n(t-t_0)$ the mean anomaly, $n=\frac{2\pi}{P}$ the mean motion with $P$ the orbital period, $\varphi_0$ and $t_0$ the initial orbital phase and initial time, respectively. In Eq. \eqref{eq:QuasiKepPar3}, the factor $2\pi/K$ gives the angle of advance of the periastron per orbital revolution, where $K$ is a function of $\nu, E, h$\footnote{The parameter $K$ is defined as $\frac{\Phi}{2\pi}-1$, see Eq. (20k) in Ref. \cite{Memmesheimer-2004}.}. Finally, $a_r$ is the semi-major axis of the orbit, $e_r, e_{\varphi}, e_t$ the PN eccentricities, and $f_{4\varphi},g_{4\varphi},f_{4t},g_{4t}$ some PN functions depending on $\nu, E, h$. The explicit expressions of all the aforementioned quantities can be found in Refs. \cite{Memmesheimer-2004,SchaferWex-1993}. 

Now, let us singularly analyse the system of Eq. \eqref{eq:QuasiKepPar}: Eq. \eqref{eq:QuasiKepPar1} is the parametrization of the orbit (where $a_r$ has been divided by $GM$ to be in agreement with the definition of $r$); Eq. \eqref{eq:QuasiKepPar2} is the 2PN version of the Kepler equation; Eq. \eqref{eq:QuasiKepPar3} gives the equation connecting the polar angle with the true anomaly, where at the 2PN order $v$ is not linearly linked to $\varphi$, as there is the addition of trigonometric functions of $v$ weighted with some PN coefficients; eventually, Eq. \eqref{eq:QuasiKepPar4} gives the classical relation existing between true and eccentric anomalies. 

\section{Elliptic case}
\label{sec:elliptic}
We assume the two bodies in the binary system move on eccentric orbits (i.e., $e_r\neq0,e_t\neq0,e_\varphi\neq0$). In this framework, we derive the analytical expression of the coordinate time $t$ in terms of the eccentric anomaly $u$. This section is organized in three parts: we first derive the function pertaining to the time that must be integrated (see Sect. \ref{sec:IF}); in this derivation, we display the presence of infinite sums and how to treat them (see Sect. \ref{sec:infinite-sums}); after these preparatory steps, we show how to derive the analytical expression of the coordinate time up to the 2PN order (see Sect. \ref{sec:analytical}).

\subsection{Integrating function}
\label{sec:IF}
The differential equation underlying the coordinate time is ruled by Eq. \eqref{eq:EoM1}. However, this form is not suitable yet, because we would obtain $t(r)$. However, from the Kepliarian-like parametrization \eqref{eq:QuasiKepPar1}, we have that $r=r(u)$. This suggests to first combine Eqs. \eqref{eq:EoM1} and \eqref{eq:EoM2} by introducing this helping function
\begin{align}\label{eq:ODE1}
    \psi (s)&\equiv\frac{\dd\varphi}{\dd t}=\frac{\dd\varphi}{\dd s}\cdot\frac{\dd s}{\dd t}\notag\\
    &=s^2\frac{\left(\sum_{j=0}^{3} B_j s^j\right)\sqrt{\sum_{j=0}^{5} A_j s^j}}{\sqrt{(s_{-}-s)(s-s_{+})}}.
\end{align}
Knowing that $s=s(u)=1/r(u)$, we arrive to
\begin{equation}\label{eq:ODE2}
    \dd t=\frac{1}{\psi (s(u))}\dd\varphi=\frac{1}{\psi (s(u))}\frac{\dd\varphi}{\dd u}\dd u\equiv \tau(u)\dd u,
\end{equation}
where $\tau(u)=\dfrac{1}{\psi (s(u))}\dfrac{\dd\varphi}{\dd u}$ and deriving Eq. \eqref{eq:QuasiKepPar3} we have
\begin{align}\label{eq:ODE3}
    \frac{\dd\varphi}{\dd u}&=\left(\frac{2\pi}{K}\right)^{-1}\frac{\sqrt{1-{e_{\varphi}}^2}}{1-e_{\varphi}\cos u}\notag\\
    &\times\Biggr[1+2\frac{f_{4\varphi}}{c^4}\cos(2v)+3\frac{g_{4\varphi}}{c^4}\cos(3v)\Biggr].
\end{align}
Equation \eqref{eq:ODE2} cannot be still integrated, because in the above expression there is $v$, which can be linked to $u$ via Eq. \eqref{eq:QuasiKepPar4}. However, this simple substitution would provide a highly nonlinear differential equation. For that reason, following the approach of Ref. \cite{Boetzel-2017}, Eq. \eqref{eq:ODE3} can be equivalently written as (see Appendix \ref{sec:Appendix1}, for details): 
\begin{subequations}\label{eq:vINu}
\begin{align}
      \cos(2v)&=\frac{2-{e_{\varphi}}^2-2\sqrt{1-{e_{\varphi}}^2}}{{e_{\varphi}}^2}+\frac{4\sqrt{1-{e_{\varphi}}^2}}{{e_{\varphi}}^2}\notag\\
      &\times\left[\sum_{m=1}^{+\infty}\beta^m \left(m\sqrt{1-{e_{\varphi}}^2}-1\right)\cos(mu)\right],\label{eq:vINu1}\\
      \cos(3v)&=\frac{3{e_{\varphi}}^2-4+(4-e_{\varphi}^2)\sqrt{1-{e_{\varphi}}^2}}{{e_{\phi}}^3}\notag\\
      &+\frac{2\sqrt{1-{e_{\varphi}}^2}}{{e_{\varphi}}^3}\left\{\sum_{m=1}^{+\infty}\beta^m 
      \left[2m^2(1-{e_{\varphi}}^2)\right.\right.\notag\\
      &\left.\left.-6m\sqrt{1-{e_{\varphi}}^2}+4-{e_{\varphi}}^2\right]\cos(mu)\right\},\label{eq:vINu2}    
\end{align}
\end{subequations}
where $\beta=(1-\sqrt{1-{e_{\varphi}}^2})/e_{\varphi}$. Therefore, to avoid to deal with nonlinear expressions, we are forced to introduce \emph{infinite sums}, which both depend directly to $\cos(mu)$ through some different coefficients.

\subsection{Compact form of the infinite sums}
\label{sec:infinite-sums}
We have seen that Eq. \eqref{eq:vINu} introduces infinite sums in Eq. \eqref{eq:ODE2}. However, this kind of representation is useless for application purposes, as they cannot be implemented in a numerical code. Therefore, we need to truncate them to a certain finite threshold $m_{\rm max}$. This value strongly depends on the input parameter values and the chosen tolerance we would like to achieve. The form of Eq. \eqref{eq:vINu} is not helpful for our objectives. Therefore, we manipulate them by introducing a more compact expression given by
\begin{align}
\chi(m_{\rm max})&=2\frac{f_{4\varphi}}{c^4}\cos(2v)+3\frac{g_{4\varphi}}{c^4}\cos(3v)\notag\\
&=\frac{1}{c^4}\Bigg{\{}
\sum _{m=1}^{m_{\rm max}} \Big[2 f_{4\varphi0} x(m)+3 g_{4\varphi} y(m)\Big]\cos(m u)\notag\\
&+2 f_{4\varphi0} x_0+3g_{4\varphi0}y_0\Bigg{\}}, \label{eq:TERM-INFINITE}
\end{align}
where $e_0=\sqrt{1+2 E_0 {J_0}^2}$ is the 0PN eccentricity, $e_1=h_0\sqrt{-E_0}$, and the other coefficients read as
\begin{subequations}
\begin{align}
f_{4\varphi0}&=\frac{e_0^2 [\nu  (19-3 \nu )+1]}{8 h_0^4},\\
g_{4\varphi0}&=\frac{e_0^3 (1-3 \nu ) \nu }{32 h_0^4},\\
x(m)&=\frac{4 e_1 \left(2 e_1 m-\sqrt{2}\right) \left(1-\sqrt{2} e_1\right)^m}{e_0^{m+2}},\\
y(m)&=\frac{ \left[2 e_1 \left(2 e_1 m^2-3 \sqrt{2} m+e_1\right)+3\right]}{2e_0\left(\sqrt{2} e_1 m-1\right)}x(m),\\
x_0&=\frac{2 e_1 \left(e_1-\sqrt{2}\right)+1}{e_0^2},\\
y_0&=\frac{e_0^3 \left[3 e_0^2+\sqrt{2} e_1 \left(2 e_1^2+3\right)-4\right]}{\left(1-2 e_1^2\right)^3}.
\end{align}    
\end{subequations}
Since the problem is invariant with respect to time- and polar-angle-shifts, we can set without loss of generality $t_0=0$ and $\varphi_0=0$. We fix the initial conditions as follows: $R_0=r(0)$ being the initial separation between the bodies; $\dot{r}(0)=0$ because we assume that the motion starts with no radial advance; $\dot{\varphi}(t_0)=\gamma\sqrt{GM/r_0^3}$, where $\gamma=\sqrt{1-e_0}$ is the initial orbit eccentricity. In this setting, the function $\chi$ depends only on $\gamma$ (or $e_0$) and $r_0$. 

\subsubsection{Analysis of the truncation error}
\label{sec:truncation}
The study of the truncation error pertaining to the $\chi$ function can be easily performed via Eq. \eqref{eq:TERM-INFINITE} through the introduction of the following function
\begin{align}
\Delta(i)=\begin{cases}
\chi(0) & \mbox{if}\ i=0,\\
\chi(i+1)-\chi(i) & \mbox{if}\ i>0,
\end{cases}
\end{align}
where $\Delta(0)$ means we are neglecting all the infinite sum, namely we have $\chi(0)=2 f_{4\varphi0} x_0+3g_{4\varphi0}y_0$. 
\begin{figure}[h!]
    \centering
    \includegraphics[scale=0.25]{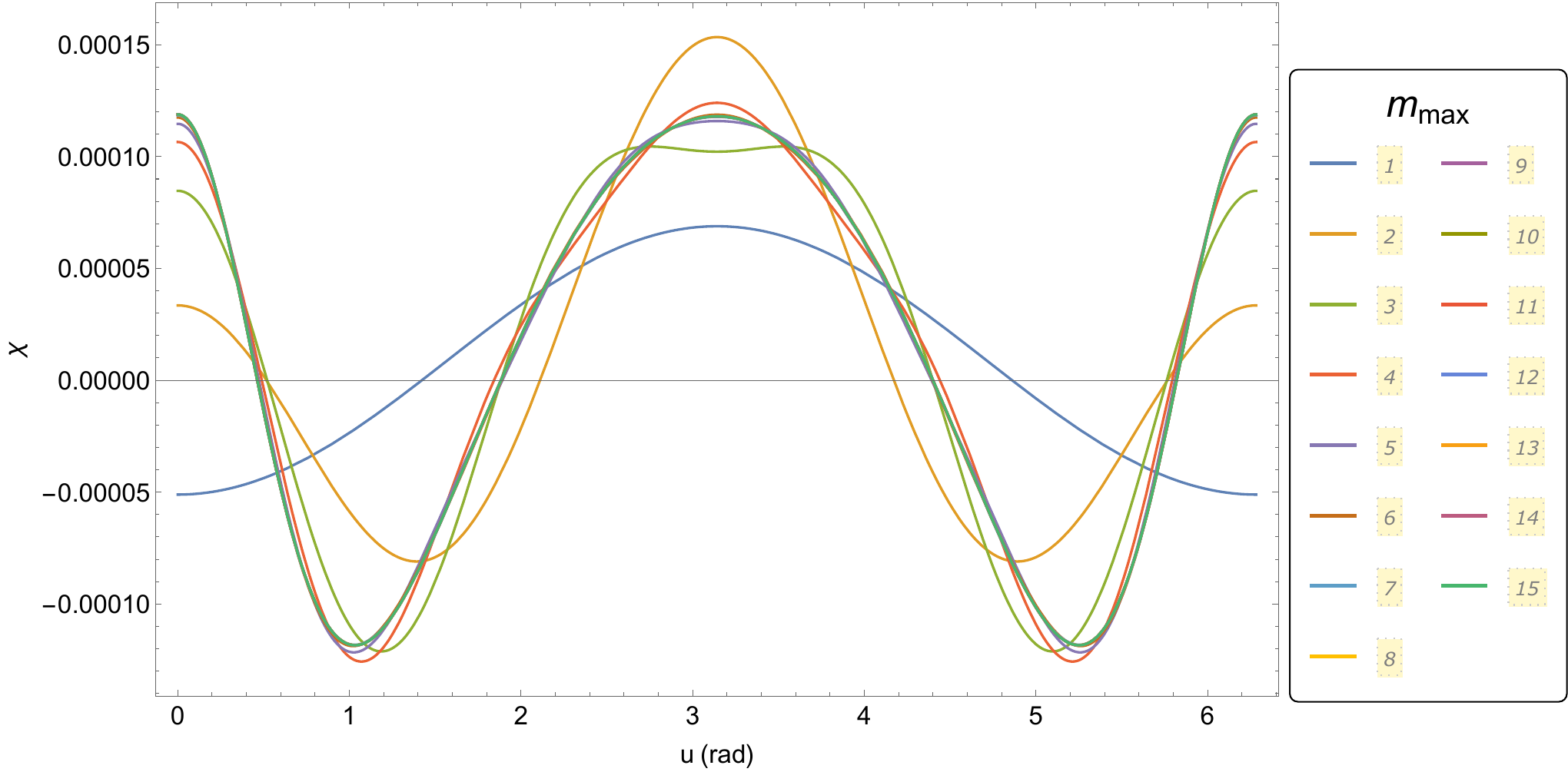}
    \caption{Plot of the function $\chi$ versus the eccentric anomaly $u\in[0,2\pi]$ for different values of $m_{\rm max}=1,\dots,15$.}
    \label{fig:Fig2}
\end{figure}
In Fig. \ref{fig:Fig2} we display the function $\chi$ versus $u$ for different values of $m_{\rm max}$. Substituting in $\chi$ these expressions $\dot{r}(0)=0$, $m_1=\frac{2}{3}M$, $m_2=\frac{1}{3}M$ and letting $r_0,\gamma,M$ free to vary, it is possible to check that $\chi$ depends only on $r_0$ and $\gamma$ (or equivalently from $e_0$). To better understand the relation existing among $\Delta,r_0,e_0$ we produce a three-dimensional plot fixing the value of $m_{\rm max}$ and evaluating $\chi$ in $u=\pi$. It is important to note that the selected value of $u$ corresponds to the maximum achieved by $\chi$, as it can be seen from Fig. \ref{fig:Fig2}.  
\begin{figure}[h!]
    \centering
    \includegraphics[trim=2cm 6cm 2cm 8cm,scale=0.28]{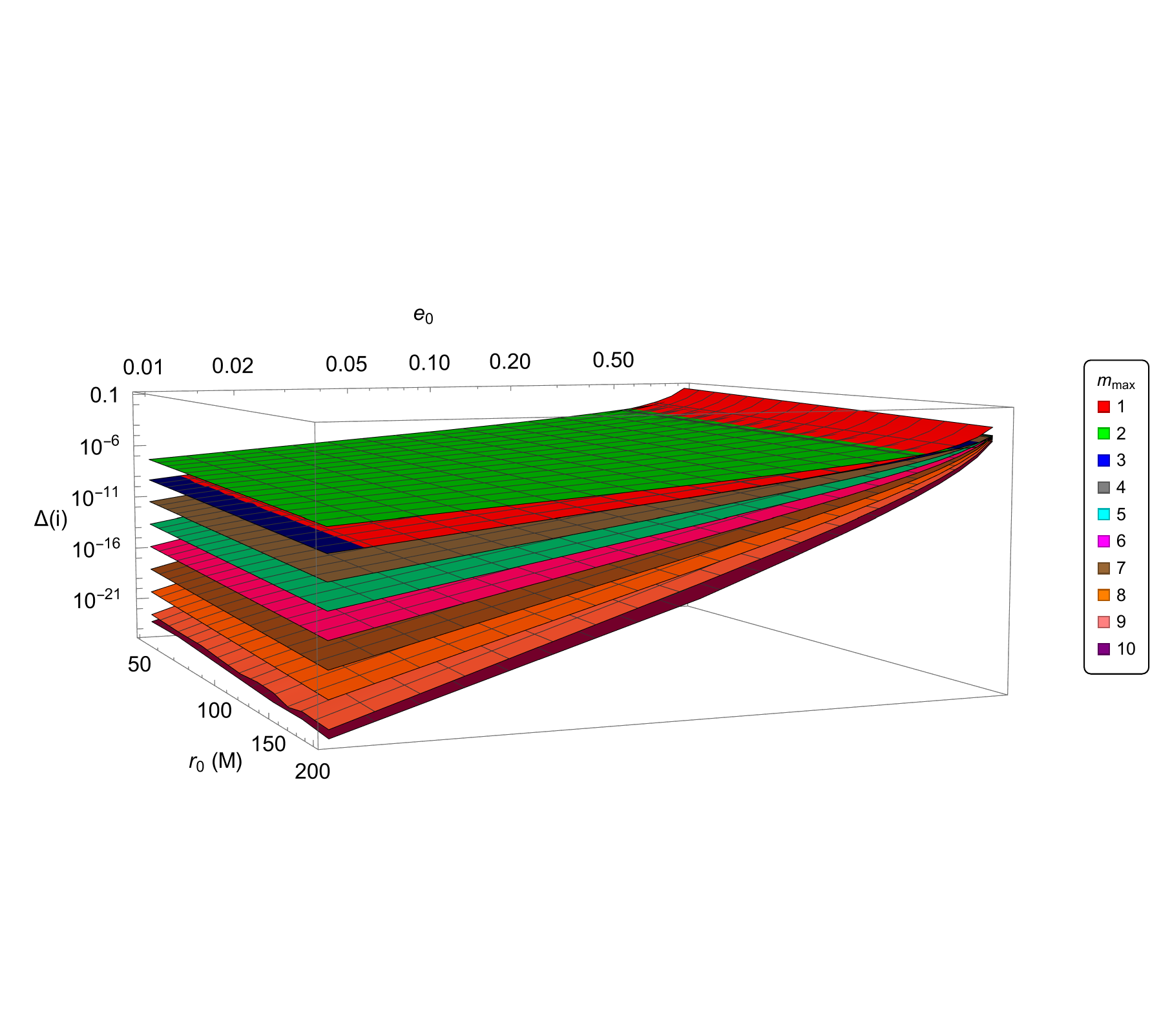}
    \caption{Three-dimensional plot showing the link among $\Delta(i),r_0,e_0$ for ten different fixed values of $m_{\rm max}$.}
    \label{fig:Fig3}
\end{figure}
In Fig. \ref{fig:Fig3} we conclude that high eccentricities $e_0$ (or equivalently low initial orbit eccentricities $\gamma$) are cause for the increment of the truncation error, whereas higher initial distances $r_0$ make it lower.

\subsection{Analytical expression}
\label{sec:analytical}
In this part, we explain how to obtain the analytical expression of the coordinate time at the 2PN order. We structure the argument in these five parts: we first write down the integrating function (see Sect. \ref{sec:general-setting}); then we easily obtain the coordinate time at 1PN order (see Sect. \ref{sec:1PN-formula}); as further step, we work out the 2PN term (see Sect. \ref{sec:2PN-formula}), which requires to introduce the concept of accumulation function to make $t(u)$ as a smooth function (see Sect. \ref{sec:accumulation}); finally, we study the accuracy of the determined analytical formula (see Sect. \ref{sec:accuracy}).

\subsubsection{General setting}
\label{sec:general-setting}
The integrating function $\tau(u)$ (cf. Eq. \eqref{eq:ODE2}) can be split at the various PN orders as follows:
\begin{equation}\label{eq:IntFun}
     \tau(u)=\tau_{\rm 0PN}(u)+\tau_{\rm 1PN}(u)+\tau_{\rm 2PN}(u).
\end{equation}
The analytical formula of the coordinate time as a function of $u$ is obtained by solving the following integrals:
\begin{align}\label{eq:Int2}
   t(u)&=\int \Big{[}\tau_{\rm 0PN}(u)+\frac{1}{c^2}\tau_{\rm 1PN}(u)+\frac{1}{c^4}\tau_{\rm 2PN}(u)\Big{]} \,\dd u\notag\\
   &=t_{\rm 0PN}(u)+\frac{1}{c^2}t_{\rm 1PN}(u)+\frac{1}{c^4}t_{\rm 2PN}(u).
\end{align}

The numerical integration of Eq. \eqref{eq:Int2} presents a monotonically increasing trend, as we expect \cite{LetteraDBA}. Instead, the analytical solution $t(u)$ can be derived by explicitly computing the three integrals in Eq. \eqref{eq:Int2}. 

\subsubsection{1PN formula}
\label{sec:1PN-formula}
The first two terms in Eq. \eqref{eq:Int2} are easily integrated, providing thus the 1PN formula of the coordinate time
\begin{subequations} \label{eq:TIME1PN}
\begin{align}
t_{\rm 0PN}(u)&=\frac{u-e_0 \sin u}{2 \sqrt{2} \left(-E_0\right)^{3/2}},\label{eq:SOL-0PN}\\
t_{\rm 1PN}(u)&=\frac{1}{8 \sqrt{2}e_0 (-E_0)^{5/2} }\biggl\{E_0^2 \biggl[8 h_0 h_1+4 e_1^2 (3 \nu -1)\notag\\
&-7 (\nu +1)\biggl]\sin u+2 e_1 \left(4 e_1^2-3\right)\sin u\notag\\
&+e_0 u \biggl[6 E_1-E_0^2 (\nu -15)\biggl]\biggl\}\label{eq:SOL-1PN}.
\end{align}
\end{subequations}
We note that this result recovers the one proposed in our previous work\footnote{We note that to compare both results, we have to consider the transformation $\varphi(u)$ or $u(\varphi)$ at 1PN level (cf. Eq. \eqref{eq:QuasiKepPar4}). In this angle conversion, the function involved is not continuous, so we must regularize it through the accumulation function.} \cite{LetteraDBA}. The main difference relies on the employment of the angle $\varphi$ inspired by the analytical expression $R(\varphi)$ provided by Damour \& Deruelle \cite{Damour1985}. However, this approach leads to discontinuous trigonometric functions. We can build up a smooth coordinate time by resorting to the accumulation function, which regularly connects the periodic branches. Here, we circumvent the aforementioned issue by exploiting the angle $u$, which directly yields a smooth function at 1PN order.   

\subsubsection{2PN formula}
\label{sec:2PN-formula}
Finally, we need to determine only the 2PN contribution $t_{\rm 2PN}$, which can be written as
\begin{align}\label{eq:IntegrandFun2PN}
  \int \tau_{\rm 2PN}(u) \,\dd u&=\mathcal{C}\left[\sum_{i=0}^{3} a_{i}\mathcal{I}(i,1)+\sum_{i=0}^{4} b_{i}\mathcal{I}(i,2)\right.\notag\\
  &\left.+\sum_{i=0}^{5} c_{i}\mathcal{I}(i,3)+\sum_{i=2}^{m_{\rm max}} d_{i}\mathcal{J}(i)\right], 
\end{align}
where the coefficients $\mathcal{C},a_{i},b_{i},c_{i},d_{i}$ depend on $\nu, E, h$ and can be found in Ref. \cite{Defalco2024}, whereas
\begin{subequations}\label{eq:Integral2PN_1}
\begin{align}
\mathcal{I}(n,m)&=\int \frac{\cos ^nu\,\dd u}{[-e_0+(1-2 e_1) \cos u] (1-e_0 \cos u)^m},\\
\mathcal{J}(n)&=\int \frac{\cos (nu)\,\dd u}{[-e_0+(1-2 e_1) \cos u] (1-e_0 \cos u)}.
\end{align}        
\end{subequations}
After the integration process, we obtain
\begin{equation}\label{eq:SOL-2PN}
    t_{\rm 2PN}(u)= D_1\mathcal{A}_1 +D_2\mathcal{A}_2 +D_3,
\end{equation}
with
\begin{subequations} \label{eq:DISC-FUNC}
\begin{align} 
\mathcal{A}_1&=\arctan{\left[\sqrt{\frac{1+e_0}{1-e_0}} \tan \left(\frac{u}{2}\right)\right]},\label{eq:coeff_SOL}\\
\mathcal{A}_2&=\arctan \left[\sqrt{\frac{e_0-2 e_1+1}{e_0+2 e_1-1}} \tan \left(\frac{u}{2}\right)\right].
\end{align}
\end{subequations}
The coefficients $D_1,D_2,D_3$ are listed in Ref. \cite{Defalco2024}. 

\subsubsection{Accumulation function}
\label{sec:accumulation}
We note that $\mathcal{A}_1$ and $\mathcal{A}_2$ have both discontinuous periodic branches. Therefore, we need to resort to the accumulation functions to obtain a smooth coordinate time. It is easy to check that $\mathcal{A}_1$ and $\mathcal{A}_2$ share both the same accumulation function, which reads as
\begin{equation} \label{eq:CONT-FUNC}
F(u)=\begin{cases}
0 & \mbox{if}\ u\in[0,\pi],\\
\pi\left\{\left[\frac{u-\pi}{2 \pi }\right]+1\right\}& \mbox{otherwise},
\end{cases}    
\end{equation}
where the symbol $[\cdot]$ in the above expression represents the integer part. Therefore, the continuous functions are
\begin{align}\label{eq:CONT}
\bar{\mathcal{A}}_1&=\mathcal{A}_1+F(u),\qquad \bar{\mathcal{A}}_2=\mathcal{A}_2+F(u).
\end{align}
Therefore, the final 2PN contribution \eqref{eq:SOL-2PN} reads as
\begin{equation}\label{eq:SOL-2PN1}
    t_{\rm 2PN}(u)= D_1\bar{\mathcal{A}}_1 +D_2\bar{\mathcal{A}}_2 +D_3.
\end{equation}

\subsubsection{Accuracy of the analytical formula}
\label{sec:accuracy}
\begin{figure}[ht!]
\includegraphics[scale = 0.3]{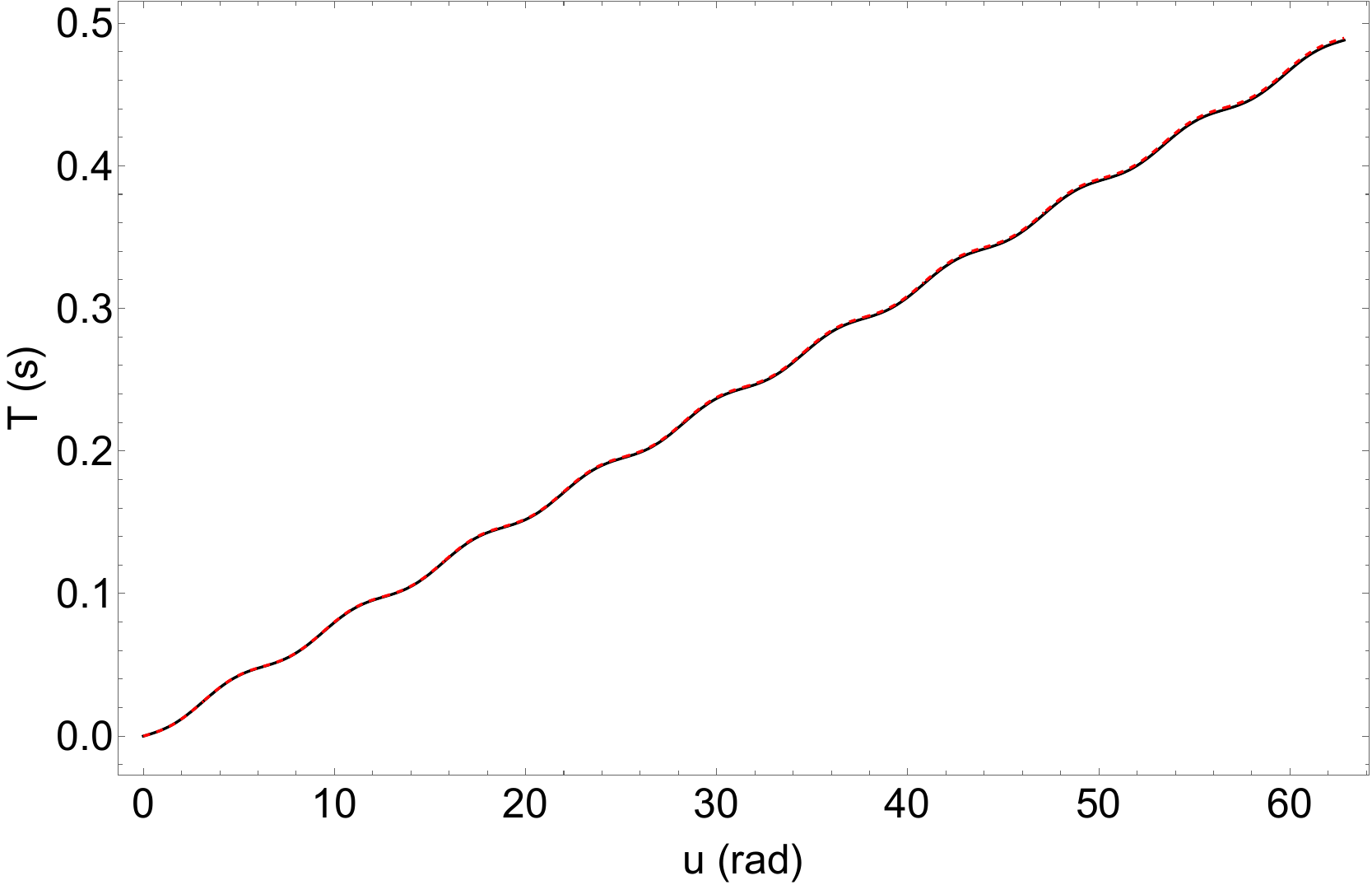}
\caption{Numerical integration of $T(u)$ (black line) and analytical formula (red dashed line) with $u\in[0,40\pi]$. The following parameter values have been used: $m_1=1.60 M_{\odot}$, $m_2=1.17 M_{\odot}$, $\gamma=0.7$, $R_0=r_0(GM)=100M$, and $\dot{r}(0)=0$. The infinite sums have been truncated at $m_{\rm max}=10$. }
\centering
\label{fig:Fig1}
\end{figure}

By substituting Eqs. \eqref{eq:SOL-0PN}, \eqref{eq:SOL-1PN}, \eqref{eq:SOL-2PN1} into Eq. \eqref{eq:Int2} we finally retrieve the analytical expression of the coordinate time at the 2PN order in terms of $u$. To check the accuracy of our analytical formula with respect to the numerical integration of Eq. \eqref{eq:Int2}, we produce the plot reported in Fig. \ref{fig:Fig1}, where we note a good agreement\footnote{In Fig. \ref{fig:Fig1}, we use the dimensional coordinate time $T$ being related to $t$ via a constant scaling term, see above Eq. \eqref{eq:EoMs} for details.}. The mean relative error, defined as the difference of the numerical and analytical expressions in modulus over the absolute value of the numerical formula, for our input parameters and selecting $m_{\rm max}=10$ amounts to $0.3\%$. We have chosen a high value of $m_{\rm max}$ just as an example to show the power of the compact form \eqref{eq:TERM-INFINITE}.

\section{Circular case}
\label{sec:circular}
In this section, we analyse the particular situation, where the two bodies move on circular orbits. This case is realized by setting 
$e_{r}=e_{t}=e_{\varphi}=0$ (see Eqs. (26a) and (26b) in Ref. \cite{Memmesheimer-2004}). This implies that the Keplerian circular orbit parametrization \eqref{eq:QuasiKepPar} becomes
\begin{align}
r=r_0,\qquad l=u=v=\frac{2\pi\varphi}{K}.   
\end{align}
It is easy to check that $f_{4t}=g_{4t}=f_{4\varphi}=g_{4\varphi}=0$ \cite{Memmesheimer-2004}.

Based on these premises, $\tau(u)$ is independent of $u$ (cf. Eq. \eqref{eq:ODE2}), as $s=1/r$ does not depend from $u$, so also $\psi(u)$ is independent from $u$. Furthermore, we have
\begin{align}
\frac{\dd \varphi}{\dd u}=\frac{K}{2\pi}.    
\end{align}
Therefore, we can write $\dd t=\mathcal{T}\dd u$, where $\mathcal{T}$ is a constant value depending only on $\nu,E,h$, which reads as: 
\begin{align}\label{eq:CircODE1}
    \mathcal{T}&=\frac{G M}{4 E_0^2 h_0}\Bigg{\{}1+\frac{2E_1 h_0^2+E_0(h_0 h_1 -3)}{c^2 e_1^2}\notag\\
    &+\frac{1}{4 c^4 e_1^4} \Bigg{[} 8e_1^2 h_0\left(E_2 h_0-E_1 h_1\right)+12E_1\left(E_1 h_0^4+2e_1^2\right)\notag\\ 
    &+4e_1^2 E_0 h_0 h_2+16e_1^6 E_0^2 \nu(2\nu+3)+2e_1^2 E_0^2 (62\nu-47)\notag\\
    &+4E_0^2 h_0 h_1 (h_0 h_1-9)-e_1^4 E_0^2 (17\nu^2 +216\nu -47)\notag\\
    &-15E_0^2(2\nu-7)\Bigg{]}\Bigg{\}}.
\end{align}
Since $t(u)=\mathcal{T} u$, the coordinate time displays a linear trend in terms of $u$, where the steepness is given by $\mathcal{T}$.

For a circular orbit, the orbital period $P$ can be computed as $P=2\pi\mathcal{T}$. It is also interesting to calculate the mean motion $\omega=\dfrac{\dd\varphi}{\dd t}=\dfrac{2\pi}{P}$, which reads as: 
\begin{align}\label{eq:CircODE2}
n&=\frac{1}{GMh_0}\Bigg{\{}-4e_1^2 E_0 + \frac{4}{c^2}\left(E_0^2 h_0 h_1-2E_1 e_1^2\right) \notag\\
&+\frac{1}{c^4}\Bigg{[} 4E_1^2 h_0^2+8E_0 E_1 h_0 h_1-8E_2 e_1^2+4E_0^2 h_0 h_2\notag\\
&+16E_0^3 e_1^4 \nu(2\nu+3)+16E_0^3(7\nu-4)-E_0^3e_1^2(17\nu^2\notag\\
&+216\nu -47)\Bigg{]}\Bigg{\}}.    
\end{align}
The case of \emph{quasi-circular orbits} (generally present in some astrophysical scenarios \cite{Alcubierre2005,Lorimer2008}) can be either computed by resorting to the coordinate time of the elliptic motion (low eccentricity, $e_0\ll1$) or can be directly approximated with that of the circular orbit case (zero eccentricity, $e_0=0$). This choice strongly relies on the input values and the accuracy we need to achieve.

\section{Astrophysical applications}
\label{sec:application}
We provide two astrophysical applications of Eq. \eqref{eq:Int2}: one related to binary NSs (see Sect. \ref{sec:aa-pulsar}) and the other pertaining to binary BHs (see Sect. \ref{sec:aa-BH}). Our formula allows us to calculate the orbital period $P$ associated with a compact binary system by assigning as input data: $m_1,m_2,R_0,e_0$\footnote{We use the dimensional initial relative radius $R_0$, being related to $r_0$ via a scaling factor, see above Eq. \eqref{eq:EoMs}.}. We note that another relevant input parameter is represented by the initial radial velocity $\dot{R}_0$, which we approximate as zero. Its value can be computed from the differential form of Kepler’s third law, since the orbital motion remains Keplerian-like at any instant \cite{DeFalco2025A}:
\begin{equation} \label{eq:rdot}
\dot{R}_0=\frac{2}{3}\frac{\dot{P}}{P}R_0,
\end{equation}
where $\dot{P}$ denotes the orbital period derivative. The compact binary systems that we consider show a $\dot{P}$ that typically lies in the range $[10^{-19},10^{-12}]$, producing $\dot{R}_0$ of the order $[10^{-10},10^{-7}]$. Such values can be reasonably assumed to be negligible, justifying thus our assumption $\dot{R}_0\approx 0$. 

The orbital separation for eccentric binaries is estimated as the apastron distance, namely $R_0=a(1+e)$, where the semi-major axis length $a$ is computed by employing again Kepler's third law yielding $a=[GM P^2/(2\pi)]^{1/3}$. These two remarks permit us to reduce the input data to only three, namely $m_1,m_2,e_0$.

We use our 2PN coordinate time formula (using $m_{\rm max}=10$) to determine the companion mass for NS binaries and the orbital period for BH binaries. In both cases, we confirm our theoretical values (denoted by an overbar, e.g., $\overline{m_2}, \overline{P}$) by benchmarking them against the observed ones reported in the literature.

\subsection{Binary neutron stars}
\label{sec:aa-pulsar}
As reported in Ref. \cite{Nair2025}, more than 30 Galactic double NS (DNS) binaries have been identified in total through radio pulsar timing, while the number of systems with reliably measured total masses amounts to about 24 of those in the Galactic field. We select six DNS sources, and we use the observational data to test Eq. \eqref{eq:Int2}, because our 2PN formula allows us to infer the companion mass $m_2$. In addition, we estimate the \emph{secular periastron advance parameter} $\dot{\omega}$ defined as $\dot{\omega}=2\pi K/P$ \cite{Kramer2021}.

Among the selected DNS systems we first consider the Hulse-Taylor gravitational source (also known as PSR B1913+16), composed of a NS and a pulsar. This is the most famous binary system, as it was not only the first ever to be discovered, but it also provided an important confirmation of GR \cite{Taylor1982}. Indeed, it has been shown that the observed orbital decay is consistent with the loss of energy due to the GW emission in GR. We choose other five candidates: PSR J0737-3039 being the first detected binary pulsar \cite{Kramer2021}, PSR J0453+1559 being another pulsar-NS asymmetric DNS having with the lightest known NS as companionn \cite{Martinez2015}, PSR J0509+3801 formed by a pulsar and NS moving on highly eccentric orbits, similarly to the Hulse-Taylor pulsar \cite{Martinez2015}, PSR J0514-4002A composed of a millisecond pulsar and a massive white dwarf (WD) companion \cite{McEwen2024,Ding2024}, and finally PSR J1518+4904 formed by a pulsar and a NS \cite{Janssen2008}.

In Table \ref{tab:DNS} we list all the input parameters related to the aforementioned DNS systems and the quantities we have calculated. It can be seen that there is good agreement between the observed values and those computed with our formula. In fact, calculating the mean relative error, it is $\sim2\times10^{-3}\%$. It is important to note that for some systems it would be possible to improve the accuracy of our predictions by considering that the bodies also are spinning, since there are relativistic spin precession effects that we do not actually take into account.
\begin{table*}[ht]
\centering
\caption{Input parameters related to the six selected DNS systems, together with the quantities estimated by us. To appreciate the accuracy and power of our formula, we consider the relative difference between the theoretical and observed values.}
\begin{tabular}{|l|ccccccccr|}
\hline
System & $m_1$  & $m_2$ & $\dfrac{|m_2-\overline{m}_2|}{m_2}$ & $R_0$ & $e_0$ & $P$ & $\dot{\omega}$ & $\dfrac{|\dot{\omega} - \overline{\dot{\omega}}|}{\dot{\omega}}$ & Ref. \\
& ($M_\odot$) & ($M_\odot$) & & ($R_\odot$) & & (d) & (deg yr$^{-1}$) & & \\
\hline
PSR B1913+16 & 1.44 & $1.39\pm0.00(1)$ & $1\times10^{-5}$ & 4.53 & 0.617 & 0.32 & 4.227 & $7\times10^{-5}$ & \cite{Weisberg2016}\\
PSR J0737$-$3039A/B & $1.34$ & $1.25 \pm 0.00(001)$ & $6\times10^{-5}$ & $1.37$ & 0.088  & 0.10 & 16.899 & $7\times10^{-4}$ & \cite{Kramer2021} \\
PSR J0453+1559 & $1.56$ & $1.17\pm0.00(4)$ & $6\times10^{-6}$ &  $16.69$ & $0.113$ & $4.07$ & 0.038 & $6\times10^{-4}$ & \cite{Martinez2015} \\
PSR J0509+3801 & $1.40$ & $1.41\pm0.00(1)$ & $1\times10^{-5}$ &  $4.94$ & $0.586$ & $0.38$ & 3.035 & $5\times10^{-4}$ & \cite{McEwen2024,Ding2024} \\
PSR J0514$-$4002A & $1.39$ & $1.08\pm0.00(1)$ & $7\times10^{-7}$ & $75.89$ & $0.888$ & $18.79$ & 0.013  & $4\times10^{-3}$ & \cite{Freire2007DNS} \\
PSR J1518+4904 & $1.47^{+0.03}_{-0.03}$ & $1.25^{+0.04}_{-0.03}$ & $3\times10^{-6}$ &  $30.88$ & $0.249$ & $8.63$ & 0.011 & $8\times10^{-4}$ & \cite{Janssen2008} \\
\hline
\end{tabular}
\label{tab:DNS}
\end{table*}

Among the $\sim3600$ pulsars identified in the Milky Way, a substantial subset resides in binary systems, most commonly with WD companions. Estimates place the Galactic population of pulsar–WD binaries at roughly 200 \cite{Lorimer2012,Tauris2023}. We apply Eq. \eqref{eq:Int2} to this class, which is dynamically less relativistic than the previously considered cases, to highlight the robustness of our formulation. In Table \ref{tab:NSWD} we report the input parameters of some pulsar-WD systems. In this case we estimate only the companion mass, because the observability of the relativistic precession advance strongly depends on the orbital eccentricity. For the majority of pulsar–WD binaries with nearly circular orbits, it is not measurable, whereas in rare eccentric systems it provides a powerful test of GR. Also in this case, we note a very good agreement between our theoretical predictions and the observational data, as the mean relative error is $\sim10^{-3}\%$. Since the eccentricities of these systems are very small, we checked our results also by employing Eq. \eqref{eq:CircODE1}, which leads to having a mean relative error of $2\%$ with respect to the eccentric 2PN formula. We stress that Eq. \eqref{eq:Int2} is too accurate for the aforementioned binary systems, as they present mild relativistic effects.

\begin{table*}[ht]
\centering
\caption{Parameters of selected NS–WD binaries together with the relative difference between companion mass values.}
\begin{tabular}{|l|ccccccr|}
\hline
System & $m_1$ & $m_2$ & $\dfrac{|m_2-\overline{m}_2|}{m_2}$ & $R_0$ & $e_0$ & $P$ & Ref. \\
& $(M_\odot)$ & $(M_\odot)$ & & $(R_\odot)$ & & (d) &\\
\hline
PSR J1141-6545 & 1.30 & $0.99\pm0.00(2)$ & $4\times10^{-5}$ & 2.20 & 0.172 & 0.20  & \cite{Bailes2003} \\
PSR B1620-26 & 1.34 & $0.34\pm0.00(4)$ & $9\times10^{-7}$ & $170.12$ & 0.025 & 191.00  & \cite{Sigurdsson2003} \\
PSR J2234+0611 & 1.35 & 0.30$^{+0.02}_{-0.01}$  & $3\times10^{-6}$ & 51.40 & 0.129 & 32.00 & \cite{Stovall2019} \\
PSR J1950+2414 & 1.50 & $0.28^{+0.01}_{-0.00(4)}$ & $4\times10^{-6}$ & 43.21 & 0.08 & 22.00 & \cite{Zhu2019} \\
PSR J0955-6150 & 1.71 & $0.25 \pm 0.002$ & $5\times10^{-6}$ & 49.90 & 0.120 & 24.58 & \cite{Serylak2022} \\
\hline
\end{tabular}
\label{tab:NSWD}
\end{table*}

\subsection{Binary black holes}
\label{sec:aa-BH}
We now consider eccentric, non-spinning BH binaries, following Ref. \cite{Ficarra2025}. That study presents a systematic numerical relativity analysis aimed at refining simulation strategies and assessing the impact of eccentricity and mass ratio on merger times and gravitational waveforms, ultimately contributing to the construction of a waveform catalog for GW astronomy. Using the \texttt{Einstein Toolkit}, the authors performed 30 simulations covering mass ratios up to $m_1/m_2=8.5$ and initial eccentricities as large as $e_0\approx0.45$. The main diagnostic is the merger time, the coordinate instant when a common horizon emerges, normalized to its quasi-circular counterpart. From these runs, they extracted the relevant parameters, summarized in Table VII of Ref. \cite{Ficarra2025}.

We select six simulations reported in Table \ref{tab:BH}. In all cases, we set the total mass of the system to $M=20\ M_\odot$. However, the subsequent analysis is independent of this number, since all quantities are scaled by the total mass of the system, as required by general covariance. The orbital period is calculated using Kepler’s third law:
\begin{equation}
P=\dfrac{2\pi}{\sqrt{\dfrac{GM}{R_0^3}}}.    
\end{equation}

As can be noted, the values estimated using our formula agree with those reported in the literature, with a mean relative accuracy of about $4\%$, further confirming the predictive power of our approach.

\begin{table}[ht]
\centering
\caption{Input parameters for BH binaries taken from Ref. \cite{Ficarra2025}, together with the estimation of the orbital period.}
\begin{tabular}{|l|cccccc|}
\hline
System & $m_1$ & $m_2$ & $R_0$ & $e_0$ & $P$ & $\dfrac{|P-\overline{P}|}{P}$ \\
& $(20 M_\odot)$ & $(20 M_\odot)$ & (M) &  & (d) & \\
\hline
\texttt{EccBBH::13} & 0.75 & 0.25 & 18.55 & 0.24 & 0.05 & $1\times10^{-2}$ \\
\texttt{EccBBH::16} & 0.80 & 0.20 & 17.76 & 0.24 & 0.05 & $3\times10^{-2}$ \\
\texttt{EccBBH::19} & 0.80 & 0.20 & 19.30 & 0.35 & 0.05 & $3\times10^{-3}$ \\
\texttt{EccBBH::23} & 0.85 & 0.15 & 16.83 & 0.24 & 0.04 & $5\times10^{-2}$ \\
\texttt{EccBBH::24} & 0.85 & 0.15 & 18.29 & 0.35 & 0.05 & $4\times10^{-2}$ \\
\texttt{EccBBH::28} & 0.90 & 0.10 & 15.69 & 0.24 & 0.04 & $9\times10^{-2}$ \\
\hline
\hline
\end{tabular}
\label{tab:BH}
\end{table}

\section{Conclusions}
\label{sec:end}
In this paper, we provide further details on the derivation of the analytical expression for the coordinate time \eqref{eq:SOL-2PN1}, previously introduced in Ref. \cite{Defalco2024}. Our guiding strategy is to follow the orbital parameterization given in Eq. \eqref{eq:QuasiKepPar}. We note that the inclusion of 2PN effects renders the classical and 1PN equations of celestial mechanics highly nonlinear. These effects are encoded in the coefficients $f_{4\varphi}, g_{4\varphi}, f_{4t}, g_{4t}$, which are accompanied by trigonometric functions of the true anomaly $v$, itself nonlinearly dependent on the eccentric anomaly $u$. This intricate structure prevents us from deriving an analytical expression for the relative separation radius $R$ as a function of $\varphi$. We only have $R(u)$, given \textit{a priori}.

The first step is to derive the differential equation governing the coordinate time, which can be written as $\dd T = f(u),\dd u$, where $f(u)$ is expanded up to 2PN order. Starting from Eq. \eqref{eq:EoM2} and expressing it in terms of the inverse radius $s$, together with the parametrization \eqref{eq:QuasiKepPar4}, we still obtain $\dd \varphi$. Therefore, calculating the Jacobian of the transformation $\dd \varphi/\dd u$, it seems that we have found the desired integrating function. However, the last term involves only $v$, which is not a simple function of $u$. A nontrivial procedure allows one to express the last term as an infinite series of trigonometric functions of $u$ (see Eq. \eqref{eq:vINu} and Appendix \ref{sec:Appendix1} for details). After some lengthy algebraic manipulations and 2PN expansions, we finally obtain the aforementioned differential equation. 

In general, the procedure mentioned above to obtain the differential equation for the coordinate time could require long calculations, but it should also lead straightforwardly to its derivation. In particular, from the 2PN order onward, it will be more and more arduous to achieve this objective for the appearance of nonlinear PN effects in the orbit parametrization. In this work, we learned how to overcome the delicate issue of the nonlinearities existing between $v$ and $u$ using the method exposed in Appendix \ref{sec:Appendix1}. In particular, we expressed such infinite sums in a more compact form through Eq. \eqref{eq:TERM-INFINITE}, which must be truncated to a certain finite value $m_{\rm max}$ for application purposes. In addition, in Sect. \ref{sec:truncation} we showed how to properly estimate $m_{\rm max}$ once the input parameters are assigned and the accuracy error is chosen.   

We stress that this preparatory part is fundamental for proceeding with the next steps, which are then expected to include a series of other demanding computations. In fact, once we obtain the function $f(u)$, we can integrate it. We note that up to the 1PN order, it does not present discontinuities, as instead they occur whether we employ the $\varphi$-parametrization and the analytical expression $R(\varphi)$ \cite{Damour1985,LetteraDBA}. However, the 2PN term presents the appearance of two tangent functions and so the occurrence of discontinuities. This issue was solved via the accumulation function to make the overall formula regular. 

This work presents two important results under a theoretical perspective in the GR PN literature: (1) the derivation of a more accurate formula for the coordinate time; (2) the refinement of a strategy from the 1PN to the 2PN order, which permits us to: gather the differential equation for the coordinate time, handle the presence of infinite sums, and to deal with discontinuous trigonometric maps via the accumulation function. These aspects have never been treated in the literature by other authors. The developed methodology can be used broadly in other research fields, which share the same starting hypotheses of the problem we studied. 

As we also underlined in the introduction, our formula is useful for computing the orbital period of compact binary systems during the inspiral stage, as well as to fit the observational data to extract information from the gravitational sources under study. We focus mainly on BH systems, as they normally feature considerable masses (and so also strong gravitational fields), which have a great impact on the 2PN term. 

In addition, we point out that we derived $T(u)$, but through Eq. \eqref{eq:QuasiKepPar3} it is also possible to obtain $T(\varphi)$. In the latter case, we should be aware that we are adding a discontinuity, so we must resort to an accumulation function. In the applications, it could happen that we need $u(T)$ or $\varphi(T)$. As the problem is formulated, we deem it is extremely complicate to determine such analytical formulae. However, we can still employ $T(u)$ or $T(\varphi)$ and invert them numerically via an interpolation procedure, which generally is not so computationally expansive.

As already stressed initially, the actual and near future upgrades of our observational capacities and sensitivities motivate us to improve the order of the PN expansions. This activity at higher and higher PN orders is fundamental not only for the practical result, but also because it allows one to figure out more about the PN effects on the coordinate time and about the existence of new advantageous parametrization, which could avoid the intervention of the accumulation function. 

Our future perspectives aim at extending the present work to the 3PN conservative dynamics and also to spinning binary systems. Regarding the last topic, we already worked along this direction, as we discovered a formal analogy existing between GR and Einstein-Cartan theory at the 1PN order \cite{Defalco2023}. Indeed, we were able to formally relate the quantum spin effects with the macroscopic angular momentum, being thus able to switch between the two cases depending on the problem under study.  

It would also be stimulating to see how this approach changes depending on 
the addition of dissipative GW effects and what kinds of new mathematical challenges will emerge \cite{Blanchet2024}. Finally, another interesting avenue would be to extend our approach to the merger and ringdown stages, where the new model of the effective-one-body must be employed \cite{Buonanno1999,Buonanno2000,Damour2011}. In this case, the technique presented in this work must be strongly refined, as the involved calculations will surely become more complicated and new fascinating issues may arise. 

\acknowledgements
VDF is grateful to Gruppo Nazionale di Fisica Matematica of Istituto Nazionale di Alta Matematica for support. VDF acknowledges the support of INFN {\it sez. di Napoli}, {\it iniziative specifiche} TEONGRAV. The authors are grateful to Alessandro Ridolfi and Delphine Perrodin for the useful discussions and suggestions. 

\appendix
\section{Derivation of infinite sums in Eq. \eqref{eq:vINu}}
\label{sec:Appendix1}
In this appendix, we outline how to derive Eq. \eqref{eq:vINu}. However, we suggest the interested reader to consult also Appendix B in Refs. \cite{Colwell1993,Boetzel-2017}, for more technical details.

Starting from $\beta=(1-\sqrt{1-{e_{\varphi}}^2})/e_{\varphi}$, we consider Eq. \eqref{eq:QuasiKepPar4}, which can be also written in terms of $\beta$ as: 
\begin{equation}\label{eq:A1_1}
    \tan{\frac{v}{2}}=\frac{1+\beta}{1-\beta}\tan{\frac{u}{2}}.
\end{equation}
Now, it is possible to show that by manipulating Eq. \eqref{eq:A1_1} by first considering the related differential equation, together with some algebraic manipulations, invoking the complex representation of trigonometric functions, and expressing the final result as convergent geometric series, we end up with the following expression \cite{Colwell1993}
\begin{equation} \label{eq:series1}
v-u=2\sum_{n=1}^\infty \frac{\beta^n}{n}\sin(nu).    
\end{equation}
One can prove that Eq. \eqref{eq:series1} can be also written as (see Appendix A in Ref. \cite{Konigsdorffer2006}, for more details)
\begin{equation}
v-u=2\arctan\left(\frac{\beta\sin u}{1-\beta\cos u}\right).     
\end{equation}
Employing the complex exponential representation of Eq. \eqref{eq:A1_1} and solving it in terms of $e^{i v}$, we obtain 
\begin{equation}\label{eq:A1_2}
    e^{iv}=\frac{e^{iu}-\beta}{1-\beta e^{iu}},
\end{equation}
where $i$ stays for the imaginary unit. Expanding then Eq. \eqref{eq:A1_2} in geometric series of $ e^{iu}$, we obtain:
\begin{equation}\label{eq:A1_3}
    e^{iv}=-\beta + \frac{2\sqrt{1-{e_{\varphi}}^2}}{e_{\varphi}} \sum_{m=1}^{+\infty}\beta^m e^{imu}.
\end{equation}
In order to have $\cos{(2v)}$ and $\cos{(3v)}$, we expand $e^{2iu}$ and $e^{3iu}$ respectively in power series, namely 
\begin{subequations}\label{eq:A1_4}
\begin{align}
      e^{2iu}&=\frac{2-{e_{\varphi}}^2-2\sqrt{1-{e_{\varphi}}^2}}{{e_{\varphi}}^2}+\frac{4\sqrt{1-{e_{\varphi}}^2}}{{e_{\varphi}}^2}\notag\\
      &\times\left[\sum_{m=1}^{+\infty}\beta^m \left(m\sqrt{1-{e_{\varphi}}^2}-1\right)e^{imu}\right],\label{eq:A1_5}\\
      e^{3iu}&=\frac{3{e_{\varphi}}^2-4+(4-e_{\varphi}^2)\sqrt{1-{e_{\varphi}}^2}}{{e_{\phi}}^3}\notag\\
      &+\frac{2\sqrt{1-{e_{\varphi}}^2}}{{e_{\varphi}}^3}\left\{\sum_{m=1}^{+\infty}\beta^m 
      \left[2m^2(1-{e_{\varphi}}^2)\right.\right.\notag\\
      &\left.\left.-6m\sqrt{1-{e_{\varphi}}^2}+4-{e_{\varphi}}^2\right]e^{imu}\right\}.\label{eq:A1_6}    
\end{align}
\end{subequations}
Finally, we recover the expressions for $\cos{(2v)}$ and $\cos{(3v)}$ as reported in Eq. \eqref{eq:vINu}, taking into account in both cases the real parts of Eq. \eqref{eq:A1_4}.

\bibliography{references}
   
\end{document}